# Design of TDC ASIC based on Temperature Compensation

*Yichao Ma[1], Xinyang Hong[1], Jian Zhuang[2,3],Zhijia Sun[2,3] , Yafan Tao[1] , Yongsheng Shi[1] , Jianrong Zhou[2,3],*
*1. Shaanxi University ofScience and Technology, Xi'an 710021, China*
*2. Institute ofHigh Energy Physics, Chinese Academy ofSciences, Beijing 100049, China*
*3. Dong guan Institute ofNeutron Science, Dong guan 523803, China*
*\* Corresponding author: Zhijia Sun , sunzj@ihep.ac.cn*

*Abstract*—.On the basis of requirement of CSNS, we designed a TDC chip with temperature compensation function in this paper, which employed TSMC 180nm process. Using delay unit bufx8 as the major method, delay lines in each level delayed input signal line through the bufx8 unit to realize fundamental measurement function. The time intervals of two fixed delay standard pulses did not change with temperature variation via intra-chip phase-locked loop. After that, the two standard pulses were sent to TDC internal delay line and measured their values. Then the measured values and standard values were compared. According to the result of comparing and decision switch, the structure of delay lines was reconstructed and their levels were recorded at the same time. We could ensure that the total length of the effective delay line were close to clock cycle as much as possible under the current temperature. The chip was tested after the completion of design. It was found that the time resolution of TDC ASIC was 73ps under 1.8V power supply at room temperature while the time resolutions were 103ps and 62ps at 85° and 0°, respectively.

**CMOS integrated circuits, Delay lines, Digital integrated circuits**

## I. INTRODUCTION

In the nuclear physics and particle physics experiments [1-4], time is one of the important physicalquantities [5-6], which measurements have important significance in nuclear electronics measurement.Through time measurement, we can obtain information on the charged particles, for example, a commonly used detector[7-9], time-of-flight detector[10-11], which can get the kinetic energy of the incoming particles by measuring the time of flight of particles. In the CSNS (Chinese spallation neutron source), we need to measure the neutron time of flight to get the neutron kinetic energy to satisfy the application scenario.

## II. ASIC DESIGN

The difficulty and emphasis in the design process of the chip is the realization of delay line (the delay line show in Fig.2). The structure of TDC is shown as Fig.1. This design adopts

TSMC 180nm process which bufx8 in the TSMC process is 53ps in the fastest delay and 113ps in the slowest delay after analysis, which all meeting the system's demand for time measurement. Therefore, we take the bufx8 as the main means to realize the delay, which delay cell is about 0.073ns in process corner ss ( ss and ff can refer to the list process corner).

For TDC satisfying application at the request of the different temperature, the delay lines promise to be long enough to meet the total delay of one clock cycle in the fastest delay time circumstance. So the total series of the delay line is 189 and the total delay time is about 10 ns in the process corner ss.

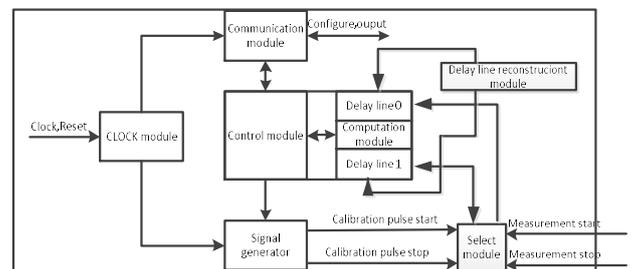

Fig.1. The Structure of The TDC Chip

The difficulty and emphasis in the design process of the chip is the realization of delay line (the delay line show in Fig.2). The structure of TDC is shown as Fig.1. This design adopts TSMC 180nm process which bufx8 in the TSMC process is 53ps in the fastest delay and 113ps in the slowest delay after analysis, which all meeting the system's demand for time measurement. Therefore, we take the bufx8 as the main means to realize the delay, which delay cell is about 0.073ns in process corner ss ( ss and ff can refer to the list process corner). For TDC satisfying application at the request of the different temperature, the delay lines promise to be long enough to meet the total delay of one clock cycle in the fastest delay time circumstance. So the total series of the delay line is 189 and the total delay time is about 10 ns in the process corner ss.

The work described in this paper was fully supported by a grant from the  the National Natural Science Foundation of China (No.11405191，11635012).

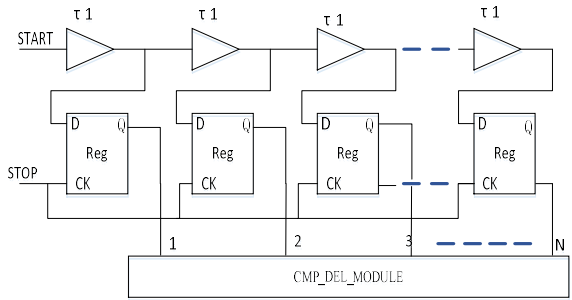

Fig.2. The Structure Diagram of Delay Line

By using the bufx8 as delay cell every level of delay line to delay the input signal line and each tap is sampled by a 100M clock. The sampling result of the delay line is output in thermometer format. Then the thermometer code is encoded by the coding module and then get the final binary delay information. The bufx8 delay in different voltage temperature is shown as Table 1.

TABLE I
BUFX8 PARAMETERS IN DIFFERENT VOLTAGE AND TEMPERATURE PARAMETERS

| porcess | Voltage(V) | Premature(℃) | Delay(ns) |
|---|---|---|---|
| tt | 1.8 | 25 | 0.072259 |
| ss | 1.98 | 0 | 0.053742 |
| ff | 1.62 | 1.25 | 0.113863ns |

## III. TEMPERATURE COMPENSATION FUNCTION TO ACHIEVE

The internal phase-locked loop chip realize two temperature-free pulse which the interval of these two pulses should be t0 (t0 only includes routing delay). The pulse as the input of the TDC internal delay line, which can obtain the measured value t1 after the delay line measurement. Compare the measured value t1 with the standard value t0, we can get the results of the comparison and according to the comparison result, combined with the judging switch the delay line structure is reconstructed and the number of delay lines series is recorded. The effective length of the delay line as close as possible in 10ns as one clock cycle with the same temperature. For example, in process corner tt, the judgment switch makes the effective delay line series a total of 136 and the series of the delay line is 136. In process corner ff, the judgment switch makes the effective delay line series a total of 189. The delay line that series is recorded in the reconstruction output the binary data which can obtain the time measurement results after update according to the time delay line series in the reconstruction. Although small change taking in minimum resolution of new delay line, delay line measurement results did not change with the temperature in the test of the reconstructed delay line.

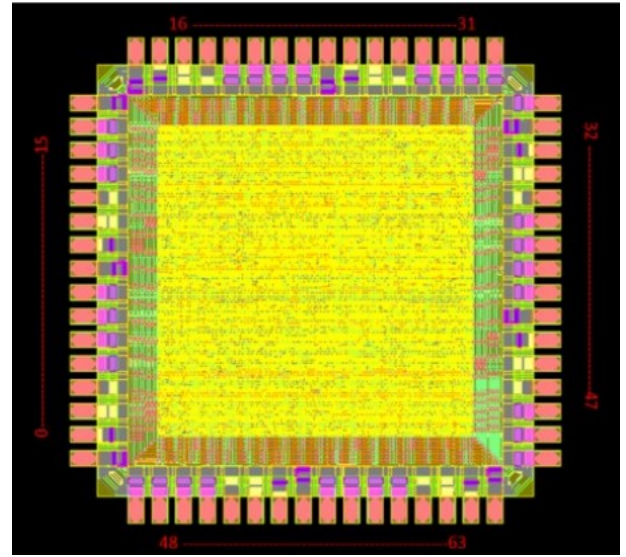

Fig.3. The TDC's Layout

The first version of the TDC ASIC contains four TDC measurement channels, which is in the trial production with the TSMC 180nm process. The Fig.3 shows the chip's layout and the Table 2 show the detail parameters of this TDC chip.

TABLE 2
THE PARAMETERS OF TDC

| area | 1643 mm X 1501mm |
|---|---|
| Static Power | 2.033e-06 W@100MHz |
| dynamic power | 0.1454+0.0269 = 0.1723W@100Mhz |
| Total Power | 0.1722W@100Mhz |

## IV. TEST RESULTS

The satisfactory result is achieved with measuring a fixed intervals pulse under differenttemperatures. The test result at different temperatures is shown as Fig4-1, Fig 4-2 , Fig4-3.

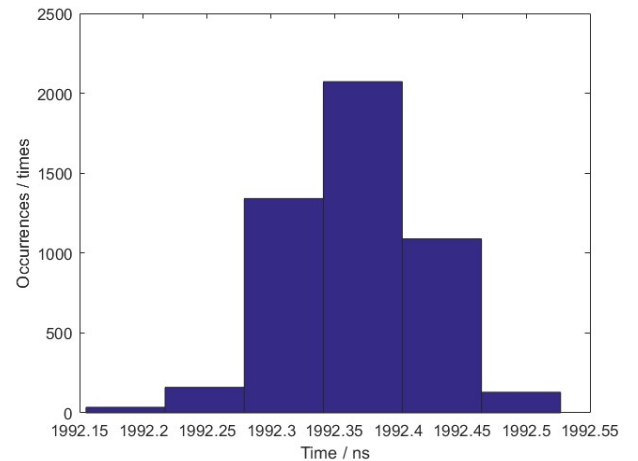

Fig.4-1. The test results in 25℃

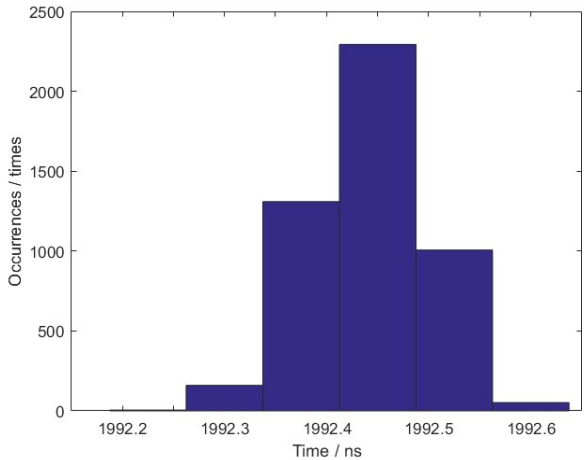

Fig.4-2 The test results in 0℃

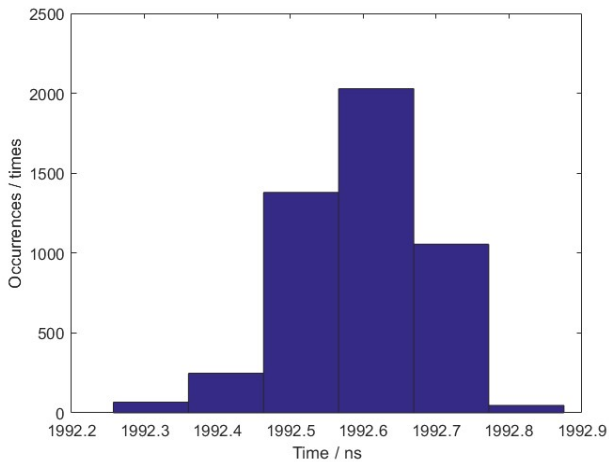

Fig.4-1 The test results in 85℃

**MaYichao,**was born in Xi'An, ShanXi in 1984. He received the B.S. and Ph.D.. degrees in modern physics from the University of Science and technology of chian, in 2011.